\documentclass[aps,twocolumn]{revtex4}
\usepackage{epsfig}
\usepackage{psfrag}
\newcommand{\be}{\begin{equation}}
\newcommand{\ee}{\end{equation}}
\newcommand{\ba}{\begin{eqnarray}}
\newcommand{\ea}{\end{eqnarray}}

\begin{document}

\title{Logarithmic corrections in (4+1)-dimensional directed percolation}

\author{Peter Grassberger}

\affiliation{John-von-Neumann Institute for Computing, Forschungszentrum J\"ulich,
D-52425 J\"ulich, Germany\\ and \\ Department of Physics and Astrophysics, University 
of Calgary, Alberta, Canada T2N 1N4}

\date{\today}

\begin{abstract}
We simulate directed site percolation on two lattices with 4 spatial and 1 
time-like dimensions (simple and body-centered hypercubic in space) with the 
standard single cluster spreading scheme. For efficiency, the code uses 
the same ingredients (hashing, histogram re-weighing, and improved
estimators) as described in Phys. Rev. {\bf E 67}, 036101 (2003).
Apart from providing the most precise estimates for $p_c$ on these lattices, 
we provide a detailed comparison with the logarithmic corrections calculated 
by Janssen and Stenull [Phys. Rev. {\bf E 69}, 016125 (2004)]. Fits with the leading 
logarithmic terms alone would give estimates of the powers of these 
logarithms which are too big by typically 50\%. When the next-to-leading 
terms are included, each of the measured quantities (the average number of sites 
wetted at time $t$, their average distance from the seed, and the probability
of cluster survival) can be fitted nearly perfectly. But these fits would not 
be mutually consistent. With a consistent set of fit parameters, one obtains 
still much improvement over the leading log - approximation. In particular 
we show that there is one combination of these three observables which seems
completely free of logarithmic terms.
\end{abstract}

\maketitle
\section{Introduction}

Although it is well known that all critical phenomena have logarithmic
corrections at their upper critical dimensions, and although the leading 
terms are easily calculated from the renormalization group, it is in 
general not easy to verify these predictions numerically. In equilibrium
models, one reason is that it is difficult to simulate a sufficiently large 
system in high dimensions, both because of storage and of CPU requirements.
The other reason is that, together with powers of the logarithm of the 
system size $L$, one usually has also terms of type $\log \log L$ etc.
If these are not known explicitly (and their computation is much more 
demanding), one has hardly any chance to verify the leading terms.

The situation is somewhat better in models with long range interactions 
\cite{luijten} and in tricritical phenomena \cite{hager} where the upper 
critical dimension is lower than in ordinary critical phenomena. It is 
also better in models like self-avoiding walks or 
percolation, where one does not need to simulate the entire lattice, but 
only fractal objects with much lower dimension. For SAWs, e.g., it was
possible to verify the structure of logarithmic corrections quite in 
detail \cite{schaefer}, since there one only has to simulate walks 
with dimension two, and since the next-to leading terms in the field
theoretic treatment could be calculated.

In the present paper we study directed percolation (DP). There, the upper 
critical dimension is 5. When interpreted as a spreading phenomenon, this 
corresponds to 4 spatial dimensions. Critical clusters then have 
spatial fractal dimension $D_f=2$, i.e. then it becomes also feasible to
study systems with very large correlation lengths. In addition, the 
leading and next-to-leading logarithmic terms have been calculated 
recently from field theory \cite{janssen}, so that we have a good 
theoretical prediction to compare with.

We study only site percolation, but on two lattices: the simple 
hypercubic (shc) lattice in 4 dimension, and the body-centered hypercubic 
(bchc) lattice. The former has $2d=8$ neighbours which can be infected in 
each time step, the latter has $2^d=16$ neighbours. We use the standard 
spreading paradigm where we start with a single infected site and 
infect in each time step neighbouring sites with probability $p$.
Sites stay infective for one time step, after that they become 
again susceptible.  We measure the average number $N(t)$ of infected 
sites, the r.m.s. distance $R(t)$ of infected sites from the seed site, 
and the probability $P(t)$ that there is still at least one infected
site (i.e., that the cluster is still alive) at time $t$. The total 
sample sizes are $5.5\times 10^7$ clusters for the shc lattice, and 
$1.5\times 10^7$ clusters for the bchc lattice, both with $t_{\rm max} = 8000$.

\begin{figure}
  \begin{center}
\psfig{file=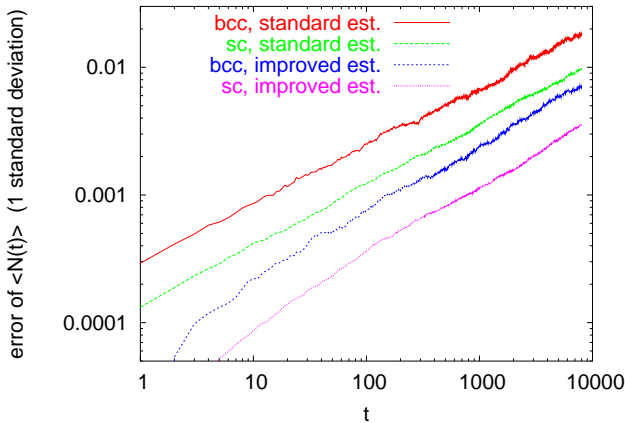,width=5.9cm, angle=270}
   \caption{(color online) Log-log plot of statistical errors (one $\sigma$) of $N(t)$ against $t$.
     The upper curves are for the usual estimates, the lower ones are for the 
     optimized improved estimates.}
   \label{1}
  \end{center}
\end{figure}

The code used to simulate this is very similar to the one used in \cite{grass}
for high-dimensional ordinary percolation:\\
1) We used hashing to store very large virtual lattices.\\
2) In addition to the straightforward averages we also estimated in each run
two averages obtained by re-weighing, corresponding to one $p-$value slightly
above and to one $p-$value slightly below the point at which we 
simulate. This is equivalent to histogram re-weighing \cite{dickman,balle},
but avoids the need for storing huge histograms.\\
3) We used improved estimators for $N(t)$ and $R(t)$, as described in 
\cite{grass}. These estimators were found to lead to large variance reduction
(the same concept was used recently also for random walks with memory,
where it also gave substantial improvements \cite{foster}). 
These estimators were found to lead to large variance reduction. 
Essentially, the idea is not to measure the actual number of offsprings in each 
generation (and their distances from the seed), but to measure the estimated
number of offsprings per active site (and their estimated distances). These 
estimates are made by counting the number of free neighbouring sites and 
multiplying it by $p$. This eliminates the fluctuations in the actual 
number of wetted sites resulting from the random number generator. Indeed,
we found that the improved estimator gave not only smaller variances than 
the standard estimator, but that the covariances between the two happened to 
be negative (we have no explanation for this lucky coincidence). Thus we can optimize the 
estimator by taking that particular linear combination which has the smallest 
variance. The resulting errors for $N(t)$ are shown in Fig.1. We see a 
reduction by roughly a factor 3, corresponding to a reduction of CPU time 
by a factor 10. The improvement was even larger (factor $\approx 4$) for 
$R(t)$. For $P(t)$ no similar improved estimator seems to exist.

\begin{figure}
  \begin{center}
\psfig{file=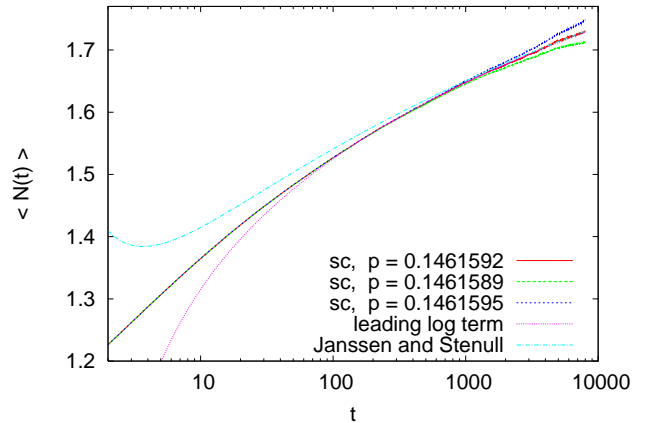,width=5.9cm, angle=270}
   \caption{(color online) Average number of infected sites, $N(t)$, for the shc lattice. The three 
    noisy curves are for $p=p_c$ and for $p=p_c\pm \Delta p_c$. The other two 
    curves show the leading log term ($\propto (\ln (t/t'_0))^{1/6}$ with $t'_0=2$)
    and the full prediction of Janssen and Stenull \cite{janssen}, Eq.~(1) with $t_0=0.5, 
    t_1 = 1.0$.}
   \label{2}
  \end{center}
\end{figure}

\begin{figure}
  \begin{center}
\psfig{file=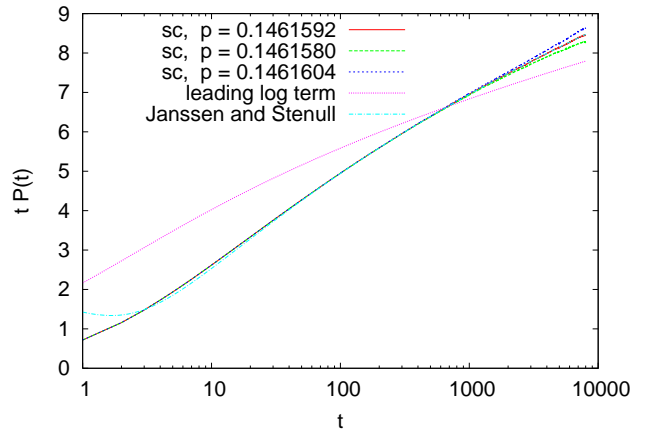,width=5.9cm, angle=270}
   \caption{(color online) Cluster survival probability multiplied by $t$, $tP(t)$, for the 
    shc lattice. The meaning of the curves is as for Fig.~2, except that the 
    leading log term is $\propto (\ln (t/t'_0))^{1/2}$ and that the three data curves 
    are separated by $4\Delta p_c$. The values for $t_0, t_1,$ and $t'_0$ are 
    the same as in Fig.~2.}
   \label{3}
  \end{center}
\end{figure}

\begin{figure}
  \begin{center}
\psfig{file=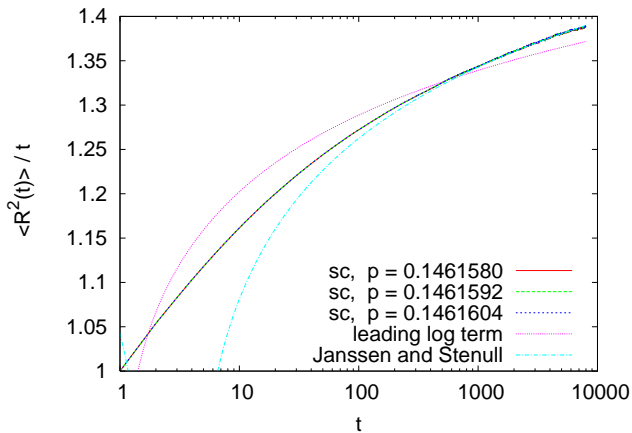,width=5.9cm, angle=270}
   \caption{(color online) Squared cluster radius divided by $t$, $R^2(t)/t$, for the
    shc lattice. The meaning of the curves is again as for Fig.~3, except that the
    leading log term is $\propto (\ln (t/t'_0))^{1/12}$.}
   \label{4}
  \end{center}
\end{figure}

\section{results}

Our main results are shown in Figs. 2 to 4. In each of them we show our 
results for the shc lattice together with the leading logarithmic term
and with a fit based on the full analytic results of \cite{janssen}. The 
integration constants $t_i$ appearing in the logarithms are the same for 
all three observables.

The results of \cite{janssen} can be rewritten as
\ba
   X_i & = & X_i^{(0)} \left[ \ln{t\over t_0} - b \ln \ln {t\over t_1} + a_i\right]^{\alpha_i} 
          \nonumber \\
   & \times &(1+O((\ln\ln t/\ln t)^2,\ln\ln t/\ln^2t,1/\ln^2t)
              \label{jans}
\ea
with $i=1,2,$ and 3. Here, 
\be
   X_1 \equiv N(t),\quad X_2 \equiv tP(t),\quad X_3 \equiv R^2(t)/t,
\ee
the exponents $\alpha_i$ are equal to 
\be
   \alpha_1 = 1/6, \quad \alpha_2 = 1/2, \quad \alpha_3 = 1/12,
\ee
the other known quantities are $ b = 1.30204,\; a_1 = 0.1831,\; a_2 = -1.5193,
\; a_3 = -1.7010,$
and $t_0$ and $t_1$ are unknown integration constants from the renormalization 
group flow. Notice that $t_0$ and $t_1$ are not universal (they differ between
models), but they are the same for all observables within one model -- although
using different values of $t_i$ for different observables could effectively 
take into account of higher order corrections.

The first observation is that the leading logarithms alone 
are not sufficient to describe the data. Using only these terms, i.e. making 
ansatzes $X_i = X_i^{(0)} [ \ln{t\over t_0}]^{\alpha_i}$, we would 
overestimate $\alpha_P$ and $\alpha_R$ by roughly 50\%. The constant $t_0$ 
can be chosen such that a nearly perfect fit is obtained for $N(t)$ at large 
$t$. But this value of $t_0$ gives bad results for the other two variables.
Also, $N(t)$ is the the variable which depends most 
sensitively on the exact value of $p_c$. It is mainly for the latter that 
we need high statistics. Without a good estimate of $p_c$ we could not 
get a decent estimate of the logarithmic corrections from the leading terms
alone. The same results were obtained for the bchc lattice (not shown here). 
Our estimates for $p_c$ are
\ba
   p_c & = & 0.0755850\pm 0.0000003 \;\;{\rm (bchc)},  \nonumber \\
   p_c & = & 0.1461592\pm 0.0000003 \;\;{\rm (shc)}.                \label{pc}
\ea

This is to be compared to Ref.~\cite{willmann}, where the authors studied 
steady state DP with
a weak rate $h$ for ``immigration" (i.e., sites are turned infective with a 
rate $h$, even when they have no infected neighbour), and then considered the 
limit $h\to 0$. The observable measured in \cite{willmann} was the density
of infected sites. Such simulations are of course much more cumbersome. In 
addition to corrections from the limit $h\to 0$ one also has finite size 
corrections which are completely absent in spreading simulations.
Indeed, the estimate for $p_c$ given in \cite{willmann}, $p_c = 0.075582\pm
0.000017$ for the bchc lattice, has an error about 60 times larger than ours.
Nevertheless, very good agreement was found in \cite{willmann} when 
comparing with the leading log terms only. We believe that this is a bit 
fortuitous.

\begin{figure}
  \begin{center}
\psfig{file=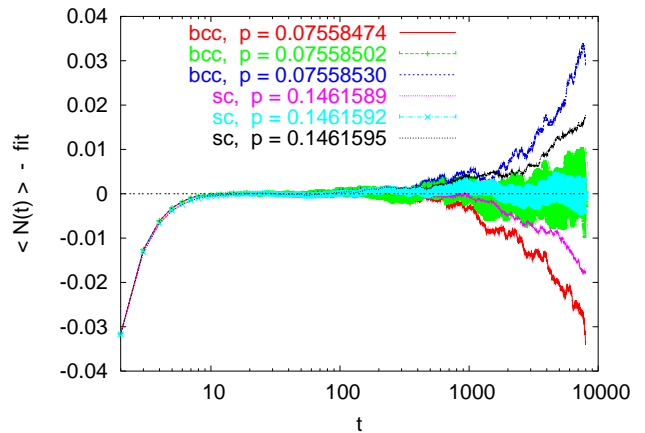, width=5.9cm, angle=270}
   \caption{(color online) Difference between the average number of infected sites $N(t)$ and 
    optimized fits with Eq.(1), both for the shc and for the bchc lattice. In the 
    fit optimal values for $t_0$ and $t_1$ are used, although these values would 
    give poor fits when used for $P(t)$ and $R(t)$.}
   \label{5}
  \end{center}
\end{figure}

The error estimates in Eq.(\ref{pc}) are of course subjective, as is true for
all extrapolations and, in particular, also for any critical exponents.
To support the above estimates 
we show in Fig.5 our values of $N(t)$, after having subtracting from 
them the best fits using Eq.(\ref{jans}). In spite of the very small error
bars of the raw data, the fits are perfect for $t>20$. The lines seen in Fig.5
correspond to $p_c \pm \Delta p_c$, with $\Delta p_c$ given in Eq.(\ref{pc}).

Unfortunately the fits used in Fig.5, although presumably correct for large values 
of $t$ and therefore suitable for estimating $p_c$, are not to be taken too 
seriously. This is seen from the fact that using 
the same values of $t_0$ and $t_1$ would give rather poor fits for the other 
two observables. As a good compromise we used $t_0=0.5,\; t_1 = 1.0$ in Figs.~2
to 4. We see that none of the three fits is perfect, but all are quite 
reasonable and definitely give a big improvement over the leading term. Thus 
we can safely conclude that the field theoretic calculations of \cite{janssen}
are verified by our simulations.

\begin{figure}
  \begin{center}
\psfig{file=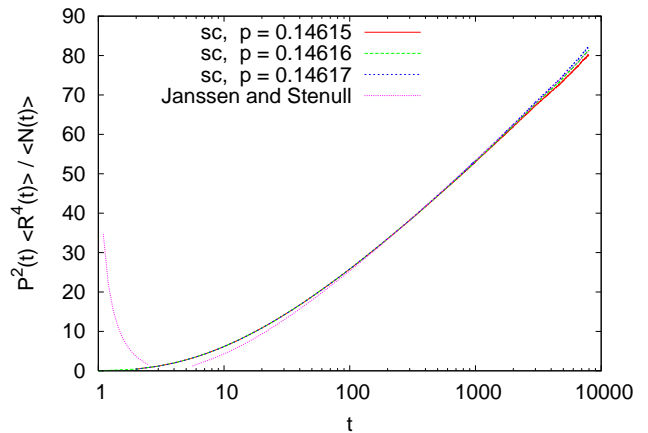, width=5.9cm, angle=270}
   \caption{(color online) Product $P^2(t)R^4(t)/N(t)$ against $t$ for the shc lattice. Notice 
    that the three curves correspond to values of $p$ whose difference is about
    30 standard deviations. The leading logarithmic correction would be linear 
    in $\ln t$.}
   \label{6}
  \end{center}
\end{figure}

\begin{figure}
  \begin{center}
\psfig{file=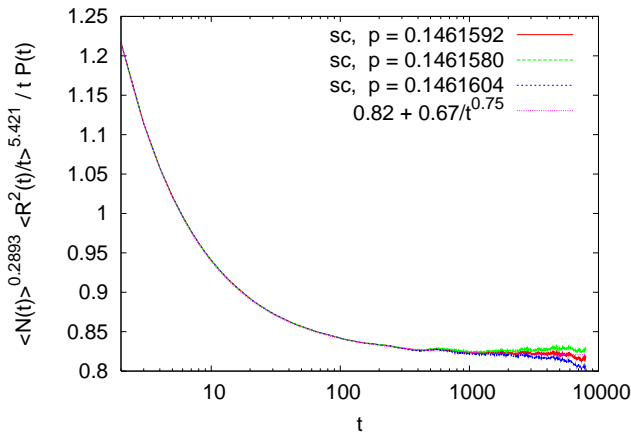, width=5.9cm, angle=270}
   \caption{(color online) Product Eq.(\ref{product}) against $t$ for the shc lattice. The 
    three data curves are for $p_c$ and for $p_c\pm 4\Delta p_c$.}
   \label{7}
  \end{center}
\end{figure}

Equation (1) was indeed obtained in Ref.[4] by first deriving parametric forms
$X_i = X_i(w)$ and $t = t(w)$, and then inverting the latter to $w=w(t)$. Since 
the parametric representations are only to lowest orders, the inversion 
introduces errors which, although subdominant asymptotically, might be 
numerically large. Comparing directly with the 
parametric expressions (Eqs.(15) and the first lines of Eqs.(25), (31), and 
(44) in Ref.~[4]) gives indeed significant further improvements for small $t$.

Before concluding, let us make two remarks. The first concerns hyperscaling.
Usually, hyperscaling is formulated in terms of critical exponents. Writing
$N(t) \sim t^\eta,\;P(t) \sim t^{-\delta},\; R^2(t) \sim t^z$ at $p=p_c$, one expects
for $d<d_c =4$ that $dz/2=2\delta+\eta$. This is no longer true for $d>4$
where $\eta=0,\delta=1,$ and $z=2$, but it still should hold in $d=4$. Written
in terms of the observables themselves, hyperscaling is equivalent (for $d<4$)
to 
\be
   P^2(t)R^d(t)/N(t) \approx const.
\ee
From Eq.(\ref{jans}) we see that this should be violated by logarithmic 
terms at $d=4$, 
\be
   P^2(t)R^d(t)/N(t) \sim [\ln t]^{2\alpha_2+2\alpha_3-\alpha_1} = \ln t.
\ee
We see from Fig.~6 that this product indeed increases strongly with $t$, 
but the increase is far from linear in $\ln t$. Thus, next-to-leading terms 
again are important. The corrections given in Eq.(1) give a big improvement,
although they are not perfect. 
An interesting observation is that this product depends 
very weakly on $p$, making it thus an ideal test object for further
non-leading logarithmic corrections. 

The second remark concerns another product of $N(t),P(t),$ and $R(t)$. 
Using Eq.(\ref{jans}) we can form one combination (and of course all its 
powers) which contains, up to the order considered in Eq.(\ref{jans}),
no logarithmic corrections at all. It is given by $\prod_i X_i^{\mu_i}$
with $\sum_i \mu_i\alpha_i = \sum_i \mu_i\alpha_i a_i = 0$. Numerically, 
we thus obtain that 
\be
   N^{0.28931}(t) \left({R(t)\over t^{1/2}}\right)^{10.8427} 
           / \left(t P(t)\right) \approx const .          \label{product}
\ee
We plot this combination in Fig.~7, together with a fit of the type 
$a + b/t^\Delta$. Numerically we found $\Delta = 0.75$. Of course one 
should not take this exponent very serious
(it could well be that the correct exponent is 1/2 or 1), but it seems
rather convincing that logarithmic terms are completely absent. Notice 
that this is not trivial. A priori, we should have expected terms 
$\sim O((\ln\ln t/\ln t)^2,\ln\ln t/\ln^2t,1/\ln^2t)$. This might hint at
a special structure of the renormalization group flow, although this 
does not seem likely from the way in which Eq.~(1) was derived \cite{janss}.

\section{Summary}

We have shown that improved algorithms for cluster spreading allow, even 
with rather modest effort
(the total CPU time used for this paper was about 1 week on a fast PC), a
rather stringent verification of logarithmic corrections at the upper critical 
dimension of one of the standard non-equilibrium critical phenomena. A 
prerequisite for this was, however, the availability of more than the 
leading log terms. If we would have had only the leading terms available
for comparison (as was the case for the steady-state equation of state studied 
in \cite{willmann}), even with much more CPU time only an order of magnitude 
verification would have been possible. 

Acknowledgments: I am indebted to Sven L\"ubeck and Richard Willmann for very 
stimulating discussions. Walter Nadler and Hannes Janssen critically 
read the manuscript, and the latter pointed out some numerical mistakes. Finally,
I want to thank an anonymous referee for suggesting an analysis directly based 
on the implicit parametric predictions of Ref.[4].

\end{document}